\documentclass[11pt]{article}
\usepackage{latexsym}  
\usepackage{amssymb}
\usepackage{amsmath}
\usepackage[dvips]{graphicx,color}

\begin{document}

\begin{flushleft}
KEK-TH-1851, KIAS-Q15004
\end{flushleft}

\begin{center}
{\Large\bf Family Gauge Boson Production at the LHC }

\vspace{4mm}
{\bf Yoshio Koide$^a$, Masato Yamanaka$^b$ and Hiroshi Yokoya$^c$ }

${}^a$ {\it Department of Physics, Osaka University, 
Toyonaka, Osaka 560-0043, Japan} \\
{\it E-mail address: koide@kuno-g.phys.sci.osaka-u.ac.jp}

${}^b$ {\it Kobayashi-Maskawa Institute for the Origin of Particles and
 the Universe (KMI),\\
 Nagoya University, Nagoya 464-8602, Japan } \\
{\it E-mail address: yamanaka@eken.phys.nagoya-u.ac.jp}

${}^c$ {\it Quantum Universe Center, KIAS, Seoul 130-722, Republic of
 Korea, \\
 Theory Center, KEK, Tsukuba 305-0801, Japan}\\
{\it E-mail address: hyokoya@kias.re.kr}

\end{center}

\vspace{3mm}

\begin{abstract}
Family gauge boson production at the LHC is investigated according 
to a $U(3)$ family gauge model with twisted family number assignment. 
In the model we study, 
a family gauge boson with the lowest mass, $A_1^{\ 1}$, interacts 
only with the first generation leptons and the third generation quarks. 
(The family numbers are assigned, for example, as 
$(e_1, e_2, e_3)= (e^-, \mu^-, \tau^-)$ and $(d_1, d_2, d_3)=(b, d, s)$
[or $(d_1, d_2, d_3)=(b, s, d)$]). 
In the model, the family gauge coupling constant is fixed by 
relating to the electroweak gauge coupling constant.  
Thus measurements of production cross sections and branching 
ratios of $A_1^{\ 1}$ clearly confirm or rule out the model. 
We calculate the cross sections of inclusive $A_1^{\ 1}$ 
production and $b \bar{b} \, (t \bar{t})$ associated $A_1^{\ 1}$ 
production at $\sqrt{s} = 14~\text{TeV}$ and $100~\text{TeV}$. 
With the dielectron production cross section, we discuss the 
determination of diagonalizing matrix of quark mass matrix, $U_{u}$ 
and $U_{d}$, respectively. 
\end{abstract}

PCAC numbers:  
  11.30.Hv, 
  12.60.-i, 
  14.70.Pw, 

\vspace{5mm}
\noindent{\large\bf 1 \ Introduction} 

One of the most challenging subjects in particle physics is to
understand the origin of ``flavor''.
It seems to be very attractive to understand ``families''
(``generations'') in quarks and leptons from concept of a
symmetry~\cite{f_symmetry}. 
Besides, in the standard model (SM) of quarks and leptons, a degree of
the freedom which is not yet accepted as a gauge symmetry is only that
of the families.
Usually, since it is considered that an energy scale of the family
symmetry breaking is extremely high (e.g.\ a GUT scale), such family
gauge models cannot be tested by terrestrial experiments. 
In addition, due to large degrees of freedoms in the models, both an
identification of each model and shedding light on model structures are
quite difficult. 
However, if a family gauge model is realized at the TeV scale and
possesses a certain degree of freedom for a clear purpose, it is worth
investigating experimental verifications of such family gauge model
seriously.

Against such conventional family gauge boson (FGB) models, recently, 
a model (Model A) with a considerably small FGB mass scale 
has been proposed by Sumino~\cite{Sumino_PLB09}.
Sumino has noticed a problem in a charged lepton mass
relation~\cite{K-mass}, 
$$ 
K \equiv \frac{m_e + m_\mu +m_\tau}{\left(\sqrt{m_e}+ \sqrt{m_\tau} 
+\sqrt{m_\tau}\right)^2} = \frac{2}{3} . 
\eqno(1.1)
$$
The relation is satisfied by the pole masses [i.e.\ $K^{pole}=(2/3) 
\times(0.999989 \pm 0.000014)$], but not so well satisfied by the
running masses [i.e.\ $K(\mu)=(2/3) \times (1.00189 \pm 0.00002)$ at
$\mu=m_Z$].
The running masses $m_{e_i}(\mu)$ are given by~\cite{Arason92}
$$
m_{e_i}(\mu) = m_{e_i} \left[ 1-\frac{\alpha_{em}(\mu)}{\pi} 
\left( 1 +\frac{3}{4} \log \frac{\mu^2}{m_{e_i}^2(\mu)} \right) \right] .
\eqno(1.2)
$$
If the family-number dependent factor $\log(m_{e_i}^2/\mu^2)$ 
in Eq.~(1.2) is absent, then the running masses $m_{e_i}(\mu)$  
also satisfy the formula (1.1). 
In order to understand this situation, Sumino has proposed a U(3) family
gauge model~\cite{Sumino_PLB09} so that a factor $\log(m_{e_i}^2/\mu^2)$
in the QED correction for the charged lepton mass $m_{e_i}$ ($i=1,2,3$)
is canceled by a factor $\log (M_{ii}^2/\mu^2)$ in a corresponding
diagram due to the FGBs.
Here, the masses of FGBs $A_i^{\ j}$, $M_{ij}$, are given by
$$
M_{ij}^2 = k (m_{e_i}^n + m_{e_j}^n),
\eqno(1.3)
$$
where $k$ is a constant with dimension of (mass)$^{2-n}$. 
The cancellation mechanism holds in any cases of $n$ in Eq.~(1.3),
because $\log M_{ii}^n =n \log M_{ii}$, although, in Model A, a case
$n=1$ has been taken. 
The cancellation condition requires the following relation between the 
family gauge coupling constant $g_F$ and QED coupling constant $e$,
$$
\left( \frac{g_F}{\sqrt2} \right)^2 =\frac{2}{n} e^2 =
\frac{4}{n} \left( \frac{g_w}{\sqrt2} \right)^2
\sin^2 \theta_w .
\eqno(1.4)
$$
Here $\theta_w$ is the Weinberg angle. 
Hence, in the FGB model we consider, the family gauge coupling constant
$g_F$ is fixed by Eq.~(1.4).

Next we see the reason why FGBs are not so heavy.  Since we consider 
$M_{ii}^2 \propto m_{e_i}^n$, the magnitude of $M_{ii}$ itself is not 
important for the cancellation mechanism, and only the linear form of 
$\log(M_{ii}/\mu)$ is essential, because $\log M_{ii}^2 = n \log 
m_{e_i} + const$. However, the cancellation 
mechanism holds only in the one-loop diagram with FGBs. 
Contributions from two loop diagrams include other forms (e.g.,\ $\log^2
M_{ii}$ and so on) in addition to the form $\log M_{ii} + const$. 
Therefore, if FGBs are too heavy, the two-loop diagrams cannot be
negligible, so that the cancellation mechanism is violated sizably. 
In other words, we cannot take the symmetry braking scale $\Lambda$ too
large. 
A speculation in Refs.~\cite{Sumino_PLB09, Sumino_JHEP09} also 
supports that the breaking scale should be intermediate scale between 
the electroweak and GUT scale. The author claims that the family gauge 
symmetry is an effective theory and must be embedded into a more 
fundamental symmetry at some scale. The scale is derived to be 
$10^{2}$-$10^{3}$\,TeV from the realization of the cancelation 
relation~(1.4) without fine tuning (details are given in 
Refs.~\cite{Sumino_PLB09, Sumino_JHEP09}).  
In this work, based on these reasons, we suppose the breaking scale 
to be $10^{3}$-$10^{4}\,\text{TeV}$. 
Let us take $n=2$ in Eq.~(1.3).
(The case with $n=2$ has been discussed in the Ref.~\cite{Koide_PLB14}.)
Then, we obtain $M_{11}/M_{33} \sim m_e/m_\tau$.
Since $m_\tau/m_e \simeq 3.5\times 10^3$, we find $M_{11} \sim$ a few 
TeV with the assumption of $M_{33} \sim \Lambda \sim 10^{4}\,\text{TeV}$.
As we show in this study, a search for a FGB with the mass $M_{11} \sim$
a few TeV is within our reach at the LHC.

An evidence of FGB can be indirectly observed from a deviation from the 
$e$-$\mu$-$\tau$ universality. 
However, such an observation has large systematic error at present, and,
for the time being, the improvement is not so easy.
Besides, even if the deviations are found, there exists various
interpretations. 
On the other hand, an observation of new vector bosons which interact
with specific family fermions can be a direct evidence at collider
experiments.
Here we note that, in general, family numbers in the lepton and quark
sectors can be assigned individually. 
As we describe in the next section, in the family gauge model we
consider, the lightest FGB couples with only first generation leptons
and third generation quarks.
Therefore, the FGB has distinguished collider signatures, such as the
characteristic production processes and their cross sections, as well as
the branching ratios. 
The complementary check by the measurements of dielectron production
cross section and $b \bar{b} \, (t \bar{t})$ associated $A_{1}^{\ 1}$
production cross sections clearly confirms or rules out the family gauge
model.
Besides, by measuring the branching ratios of the FGB, we can
distinguish whether the signal is from the FGB or not. 

This work is organized as follows.
First we briefly review our model (Model B) which is an extended version
of Model A, in particular, the interactions relevant for the collider
phenomenology.
Then we comment on the assignment of family number in quark sector with
taking into account the observational results of pseudo scalar
oscillations. 
Next, in Sec.~3, we check current direct bound on $A_1^{\ 1}$ 
from the data of LHC 8 TeV run.
Then, we evaluate the production cross section of $A_1^{\ 1}$ at
$\sqrt{s} = 14~\text{TeV}$ and $100~\text{TeV}$, and discuss the
feasibility of the discovery in future experiments.
Finally, Sec.~4 is devoted to summary and discussion.

\vspace{5mm}
\noindent{\large\bf 2 \ Family gauge boson model}

We describe the interactions and flavor structures in the family 
gauge model proposed in Ref.~\cite{Koide_PLB14}, which we call Model B, 
to discuss the collider signatures of the FGBs in the model. 
Model B is an extended model of the family gauge
model proposed in Ref.~\cite{Sumino_PLB09}, which we call Model A.
Model B improves the shortcomings of Model A, and as a 
result, characteristic interactions for the FGBs are introduced. 
We give a brief review on Model B in this section. 

In Model A \cite{Sumino_PLB09}, 
Sumino has assigned the fermions (quarks and leptons) 
$(f_L, f_R)$ to $({\bf 3}, {\bf 3}^*)$ of U(3), the gauge group of the
family gauge symmetry, in order to obtain the minus sign for the
cancellation in lepton running masses, i.e., the cancellation between a
factor $\log m_{e_i}^2$ and $\log M_{ii}^2$ (see Introduction).
Although the assignment successfully brings the cancellation 
mechanism, 
it has a shortcoming from the phenomenological point of view: 
The assignment induces effective quark-quark interactions with $\Delta
N_{fam}=2$ ($N_{fam}$ is family number).
It causes a serious conflict with the observed $P^0$-$\bar{P}^0$
mixings ($P=K, D, B, B_s$) .
Therefore, in the Model B, only for quark sector, we restore the
Sumino's assignment  $(q_L, q_R) \sim ({\bf 3}, {\bf 3}^*)$ of U(3) to
the normal assignment $(q_L, q_R) \sim ({\bf 3}, {\bf 3})$, in order to
suppress the unwelcome quark-quark interactions with $\Delta N_{fam}=2$.
We note that, in the quark sector, we do not need such cancellation
mechanism as in the charged lepton masses. 
On the other hand, the idea of cancellation mechanism is inherited in
the lepton sector from Model A.
Thus, we adopt $(\ell_L, \ell_R)\sim({\bf 3}, {\bf
3}^*)$~\cite{Sumino_PLB09} for the lepton sector, so that the relations
(1.3) and (1.4) hold for the cancellation mechanism.
As a result, the FGB interactions with quarks and leptons
are given as follows:
$$
{\cal H}_{fam} = \frac{g_F}{\sqrt{2}} \left[ \sum_{\ell= \nu, e} 
 \left( \bar{\ell}_L^i \gamma_\mu \ell_{Lj} - \bar{\ell}_{Rj} \gamma_\mu
 \ell_R^i \right) +  \sum_{q=u,d} (U_q^{*})_{i k} (U_q)_{j l}
 (\bar{q}_{k} \gamma_\mu  q_{l}) \right] (A_i^{\ j})^\mu .
\eqno(2.1)
$$
We note that the U(3) assignment for fermions is not anomaly free in
both Models A and B.
In order to avoid this shortcoming, we tacitly assume an existence of
heavy leptons in the lepton sector.

Furthermore, in the present paper, we discuss the case $n=2$, only the
case which can give $M_{11} \sim 1$ TeV~\cite{Koide_PLB14}. 
Then, the value of $g_F$ is fixed as follows:
$$
\left. \frac{g_F}{\sqrt{2}}\right|_{n=2} = e=0.30684.
\eqno(2.2)
$$

In Model B as well as in Model A,
the FGB mass matrix is diagonal in a flavor basis 
in which the charged lepton mass matrix $M_e$ is diagonal. 
There is no family number violation at the tree level in the lepton 
sector. 
However, in general, there is a mixing between the family number basis
and the mass basis in the quark sector. 
In Eq.~(2.1), $U_{q}$ is the diagonalizing matrix for the quark mass
matrix $M_q$, and the Cabibbo-Kobayashi-Maskawa (CKM)~\cite{CKM}
quark mixing matrix $V_{CKM}$ is given by $V_{CKM}=U_{Lu}^\dagger
U_{Ld}$. 
In this paper, for convenience, we assume the mass matrix is Hermitian,
thus $U_{Lq}=U_{Rq}=U_q$. 
In addition, for numerical estimates, we use an assumption, 
$$
U_u \simeq {\bf 1} , \ \ \ U_d \simeq V_{CKM} ,
\eqno(2.3)
$$
by considering the observed fact $m_t-m_u \gg m_b-m_d$.

Even if we adopt $(q_L,q_R)\sim({\bf 3},{\bf 3})$, the observed
$K^0$-$\bar{K}^0$ mixing still puts a severe constraint on the masses
$M_{ij}$.
To avoid this constraint, we can take a lower mass only for the FGB
which interacts with the third generation quarks. 
Hence, a twisted family number assignment, 
$$
(d_1, d_2, d_3)=(b, d, s) \ \ [{\rm or}  \ 
(d_1, d_2, d_3)=(b, s, d)] \ \  
  {\rm vs.}  \ \ (e_1, e_2, e_3)=(e^-, \mu^-, \tau^-)
\eqno(2.4)
$$
has been proposed for the quark sector~\cite{Koide_PLB14}.
In this case, the lightest FGB $A_1^{\ 1}$ interacts with the first
generation leptons and the third generation quarks.
Thereby, we can safely construct a family gauge model with lower mass
scale without conflicting with constraints from the observed
$P^0$-$\bar{P}^0$ mixing~\cite{Koide_PLB14}. 
This assignment (2.4) is a key idea to make the FGB model viable at the
TeV scale.

\vspace{5mm}

\noindent{\large\bf 3 \ $A_1^{\ 1}$ production at the LHC}

One of the clear observable at collider experiments in the present model
is the dielectron resonance via the lightest FGB $A_1^{\ 1}$. 
The interactions~(2.1) indicate that additional important observable is
the $A_1^{\ 1}$ production associated with third generation quarks. 
$A_1^{\ 1}$ mass is obtained by the peak position in the dielectron
invariant mass, and $A_1^{\ 1}$ interactions are, in principle,
determined by the measurement of the cross sections for these processes.
In this section, we evaluate the cross sections at the LHC, and discuss 
the feasibility of the discovery of $A_1^{\ 1}$ FGB in future collider
experiments.

\vspace{3mm}

{\it 3.1 \ Branching ratios of the FGB $A_1^{\ 1}$}

Prior to calculation of the production rate of $A_{1}^{\ 1}$ at the LHC, 
we discuss the decay rates of $A_{1}^{\ 1}$. 
Major decay modes of $A_1^{\ 1}$ are $t\bar{t}$, 
$b\bar{b}$, $e^+ e^-$ and $\nu\bar{\nu}$, where
$\nu\bar\nu$ indicates the sum of the neutrino anti-neutrino pair over
the three mass eigenstates. 
The partial decay width $\Gamma(A_1^{\ 1} \rightarrow f+\bar{f})$ 
is given by
$$
\Gamma(A_1^{\ 1} \rightarrow f+\bar{f}) = 
\frac{C}{12 \pi} \frac{g_F^2}{2} M_{11} 
\left( 1 +\frac{2 m_f^2}{ M_{11}^2} \right)
\sqrt{ 1 -\frac{4 m_f^2}{ M_{11}^2}  }  ,
\eqno(3.1)
$$
where $C$ is a factor, $C=1$ for charged leptons and 
$C=3\times |(U_q)_{1i}|^2$ for the quark pair $\bar q_i q_i$.
For neutrinos, $C=1/2$ for Majorana case, while $C=1$ for Dirac case.

$A_{1}^{\ 1}$ with its mass lighter than 1~TeV has been already
excluded by the direct search at the LHC 8~TeV run (see next subsection). 
For $M_{11} > 1$~TeV, we can approximate 
$$
\left( 1 + 2 m_f^2/ M_{11}^2 \right) 
\sqrt{  1- 4 m_f^2/M_{11}^2  } \simeq 1.
$$
With this approximation, the total decay width for $M_{11} > 1$~TeV is 
given for the case of Majorana neutrinos by
$$
\Gamma_{A_{11}} = \frac{15}{2} M_{11} [{\rm TeV}]\times 2.497 \times
10^{-3},
\eqno(3.2)
$$
and the branching ratios are as follows:
$$
\begin{array}{l}
Br( A_1^{\ 1} \rightarrow t \bar{t}) \simeq 
Br( A_1^{\ 1} \rightarrow b \bar{b}) \simeq 40 \% , \\[2mm]  
Br( A_1^{\ 1} \rightarrow e^- e^+) = \dfrac{2}{15} = 13.3 \% , \\[2mm] 
Br( A_1^{\ 1} \rightarrow \nu \bar{\nu}) = \dfrac{1}{15} = 6.7 \% , 
\end{array}
\eqno(3.3)
$$ 
while the other decay modes are zero or highly suppressed as long as we
employ the naive assumption for the quark mixing matrices in Eq.~(2.3).

In the case of Dirac neutrinos, the total decay width is given by
$\Gamma_{A_{11}} = 8M_{11} [{\rm TeV}]\times 2.497 \times
10^{-3}$, and the branching ratios are 
$$
\begin{array}{l}
Br( A_1^{\ 1} \rightarrow t\bar{t}) \simeq 
Br( A_1^{\ 1} \rightarrow b\bar{b}) \simeq 37.5 \% , \\[2mm]  
Br( A_1^{\ 1} \rightarrow e^-e^+) =
Br( A_1^{\ 1} \rightarrow \nu\bar{\nu}) = \dfrac{1}{8} = 12.5 \%,
\end{array}
\eqno(3.4)
$$
with zero or negligibly small ratios for the other modes.
The difference between the two cases is due to the number of light
neutrinos, since FGBs couple to both the left-handed and right-handed
neutrinos. 
In future, when data of the $A_1^{\ 1}$ production is accumulated, we
are able to conclude whether neutrinos are Dirac-type or Majorana-type
by measuring the branching ratios of FGBs.
Especially, the branching ratio for the invisible decay mode has the
largest difference between the two cases.

One of the ingredients to discriminate our scenario and other models is
the ratio between branching ratios of $e^+ e^-$ and $b \bar b$ (or $t
\bar t$) final states, $\text{Br} (A_1^{\ 1} \to b \bar b (t \bar
t))/\text{Br} (A_1^{\ 1} \to e^+ e^-) \simeq 3$.
This is because that various models possess an extra neutral current,
and each model predicts different partial width for each final states
$e^+ e^-$, $b \bar b$ (or $t \bar t$), and so on. We see two examples.
One of the natural classes of models that predict an extra neutral
current is extra dimension models. 
An example is the Randall-Sundrum model~\cite{Randall:1999ee}. 
The Kaluza-Klein (KK) partner of graviton $G_\text{KK}$ is a $Z'$ boson
like particle, which can produce di-top and di-electron signals. 
In this model, the ratio is $\text{Br} (G_\text{KK} \to t \bar
t)/\text{Br} (G_\text{KK} \to e^+ e^-) \gtrsim
10^{3}$~\cite{Fitzpatrick:2007qr}. 
Another example is the universal extra dimension (UED) model 
\cite{Appelquist:2000nn}. 
The second KK partner of $U(1)_Y$ gauge boson $B^{(2)}$ is also a $Z'$
boson like particle, and decays into $e^{+} e^{-}$, $b \bar b$ and
others. 
In the UED model, the ratio is $\text{Br} (B^{(2)} \to b \bar
b)/\text{Br} (B^{(2)} \to e^+ e^-) \simeq (7 - 10)$~\cite{Datta:2005zs,
Matsumoto:2009tb}.

\vspace{5mm}

{\it 3.2 \ Production rate and discovery significance}

\begin{figure}[t!]
\begin{center}
\includegraphics[width=110mm]{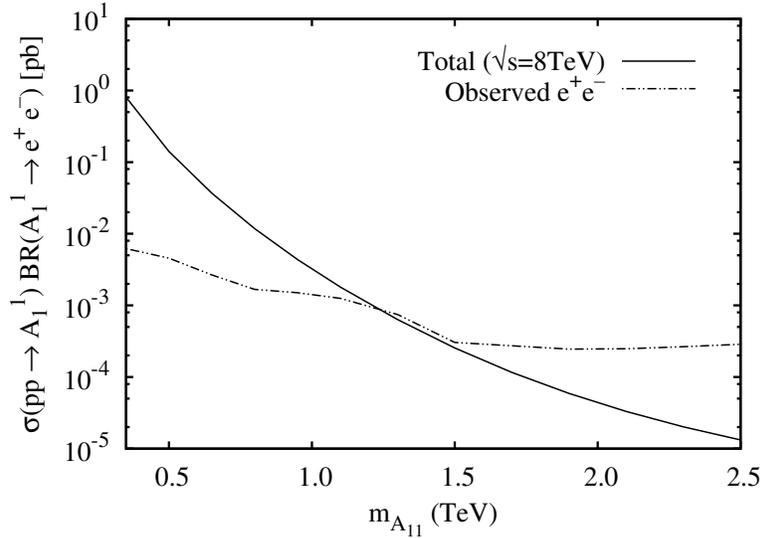}
\caption{The cross section of dielectron production via 
$A_{1}^{\ 1}$ at $\sqrt{s}=8$~TeV [Solid curve]. 
Horizontal two-dot chain curve represents the observed 
95\% C.L.\ upper limit of the cross section of dielectron 
resonance~\cite{Aad:2014cka}. }
\label{Fig:8TeV_with}
\end{center}  
\end{figure}

\begin{figure}[t!]
\begin{center}
\includegraphics[width=115mm]{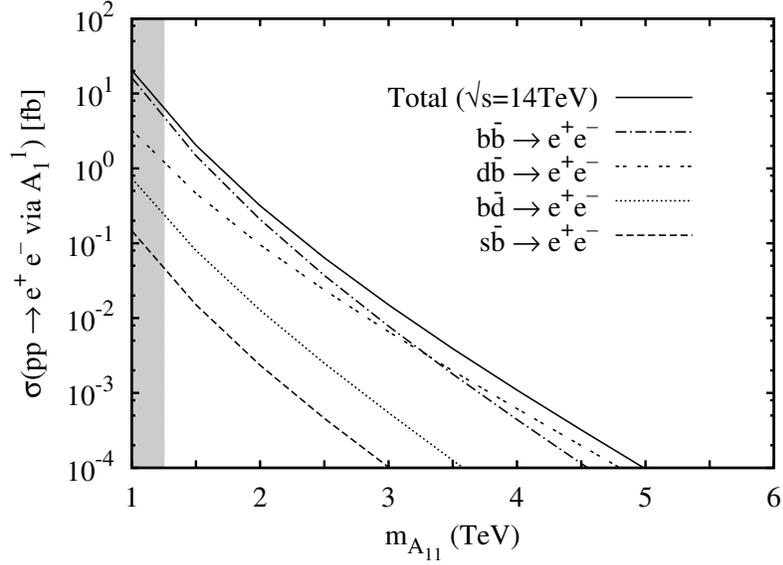}
\caption{The cross section of dielectron production via $A_{1}^{\ 1}$ at
 $\sqrt{s}=14$~TeV [Solid curve]. 
Other curves show partial cross sections for the subprocesses with large
 contributions. 
A cut on the invariant mass of the $e^{+} e^{-}$ pair is placed as
$M_{11} - 1.5 \Gamma_{A_{11}} \leq m_{ee} \leq M_{11} 
+ 1.5 \Gamma_{A_{11}}$. Shaded region is ruled out by 
dilepton search at $\sqrt{s}= 8~\text{TeV}$ LHC.}
\label{Fig:14TeV_with}
\end{center}  
\end{figure}

\begin{figure}[t!]
\begin{center}
\includegraphics[width=115mm]{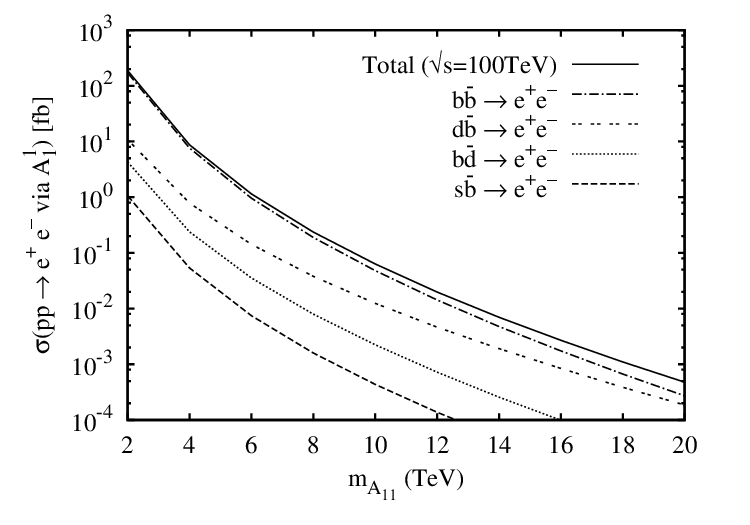}
\caption{Same as Fig.~\ref{Fig:14TeV_with}, but for the 
collision energy $\sqrt{s}=100$~TeV.}
\label{Fig:100TeV_with}
\end{center}  
\end{figure}

\begin{figure}[t!]
\begin{center}
\includegraphics[width=115mm]{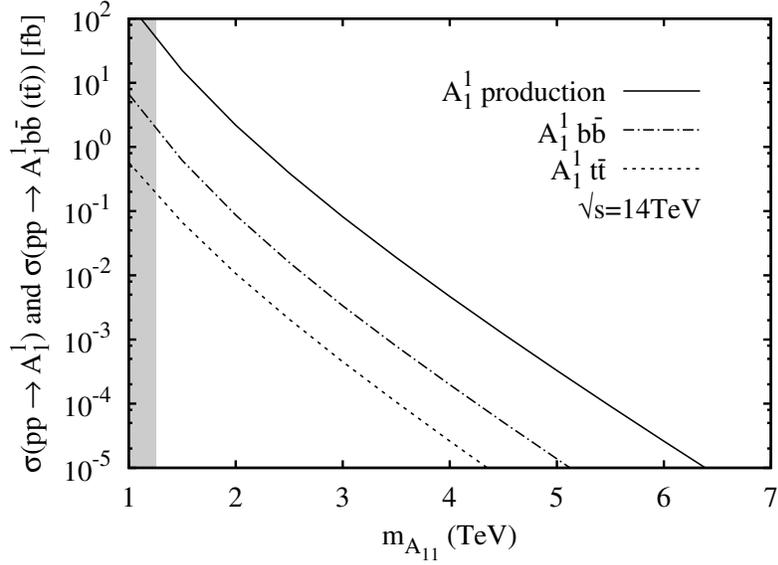}
\caption{The cross sections of $A_{1}^{\ 1}$ direct production and 
$b \bar{b} \ (t \bar{t})$ associated productions at $\sqrt{s}=14$~TeV. 
For the $b \bar{b}$ associated production, two cuts are placed: 
(i) large transverse momentum $p_{T}^{b(\bar{b})} > 25\text{GeV}$ 
(i\hspace{-1pt}i) pseudo-rapidity smaller than $|\eta| < 2.5$. Shaded 
region is ruled out by the dilepton search at $\sqrt{s}= 8~\text{TeV}$ 
LHC.}
\label{Fig:14TeV_1body_bb_tt}
\end{center}  
\end{figure}

\begin{figure}[t!]
\begin{center}
\includegraphics[width=115mm]{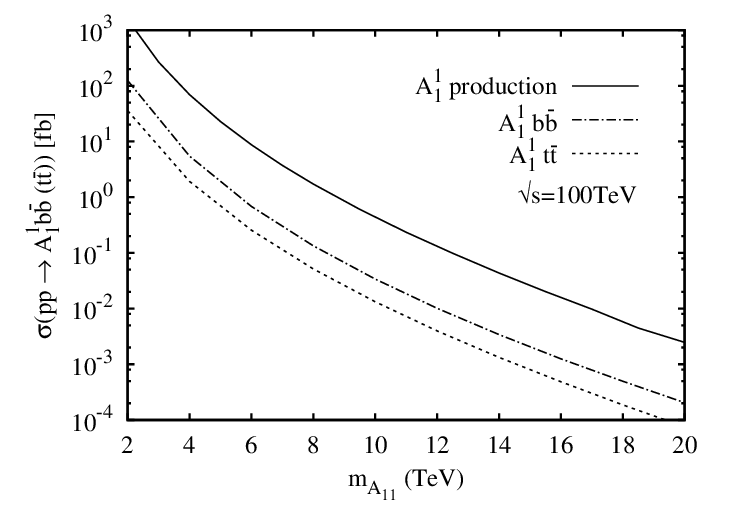}
\caption{Same as Fig.~\ref{Fig:14TeV_1body_bb_tt}, but for the 
collision energy $\sqrt{s}=100$~TeV.}
\label{Fig:100TeV_1body_bb_tt}
\end{center}  
\end{figure}

\begin{figure}[t!]
\begin{center}
\includegraphics[width=110mm]{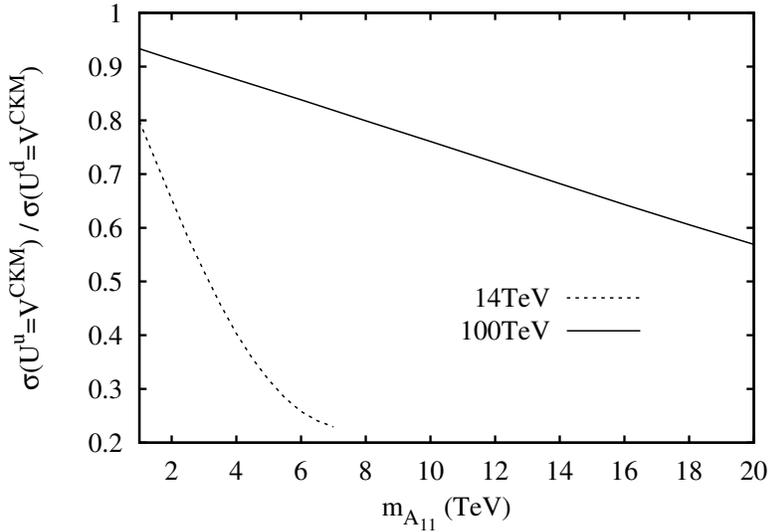}
\caption{The ratio of the dielectron production cross sections via 
$A_{1}^{\ 1}$ in the two cases:
(i) $U_{u} = (V_\text{CKM})^{\dagger}$,  
$U_{d} = \boldsymbol{1}$ and (i\hspace{-1pt}i) $U_{u} = \boldsymbol{1}$,
 $U_{d} = V^\text{CKM}$, 
 $\sigma(U^{u}=V^\text{CKM})/\sigma (U^{d}=V^\text{CKM})$.}
\label{Fig:ratio}
\end{center}  
\end{figure}

Now we evaluate the cross sections of $A_1^{\ 1}$ production, 
and see the perspective of $A_1^{\ 1}$ direct search at
the LHC and future hadron collider experiments. 
For a reference scenario, first, we take $U_{u} = \boldsymbol{1}$ and 
$U_{d} = V_\text{CKM}$.
The quark mixings in the following calculation are $|V_{td}| = 0.00886$,
$|V_{ts}| = 0.0405$, and $|V_{tb}| = 0.99914$~\cite{PDG14}. 

Before we discuss the perspective, we check the current bound on $A_1^{\
1}$ from the LHC data.
Figure~\ref{Fig:8TeV_with} shows the dielectron production cross section
via $A_1^{\ 1}$ at $\sqrt{s} = 8~\text{TeV}$ as a function of the
$A_1^{\ 1}$ mass. The cross section is calculated using
calcHEP~\cite{Belyaev:2012qa} with the CTEQ6L parton distribution
functions~\cite{Lai:1999wy}.
In the evaluation of $\sigma(pp \to A_{1}^{\
1} \to e^{+} e^{-})$, we apply a cut on the invariant mass of the $e^{+}
e^{-}$ pair, $M_{11} - 1.5 \Gamma_{A_{11}} \leq m_{ee} \leq M_{11} + 1.5
\Gamma_{A_{11}}$.
The horizontal curve represents observed 95\% C.L.\ upper limit of the
cross section of dielectron resonance with an integrated luminosity
$20.3~\text{fb}^{-1}$~\cite{Aad:2014cka}. 
By a comparison of the observed limit and our calculation, the mass
limit of $M_{11} \gtrsim 1.25~\text{TeV}$ is obtained. 
This is the current lower limit of $M_{11}$ in the scenario with $U_{u}
= \boldsymbol{1}$ and $U_{d} = V_\text{CKM}$.

\begin{table}[h]
\begin{center}
\caption{Significance of the $A_{1}^{\ 1}$ discovery on 14~TeV 
LHC run, $S/\sqrt{S+B}$, for integrated luminosity $\mathcal{L} 
= 300~\text{fb}^{-1}$ and $3000~\text{fb}^{-1}$. Here $S$ 
and $B$ are numbers of signal event and background event, 
respectively.}
\vspace{3mm}
\begin{tabular}{llllll}
\hline
$M_{11}$~[TeV]
& $\sigma_\text{BG}$~[pb] \hspace{15mm}
& $\frac{S}{\sqrt{S+B}} \ (\text{for }300~\text{fb}^{-1})$ 
& $\frac{S}{\sqrt{S+B}} \ (\text{for }3000~\text{fb}^{-1})$
\\ \hline \hline
$2.0$
& $ 6.801 \times 10^{-5}$
& $ 6.859$
& $ 21.69$
\\[0.5mm]
$2.5$
& $ 2.084 \times 10^{-5}$
& $ 2.943$
& $ 9.306$
\\[0.5mm]
$3.0$
& $ 7.072 \times 10^{-6}$
& $ 1.356$
& $ 4.287$
\\[0.5mm]
$3.5$
& $ 2.556 \times 10^{-6}$
& $ 0.653$
& $ 2.063$
\\[0.5mm]
$4.0$
& $ 9.580 \times 10^{-7}$
& $ 0.324$
& $ 1.025$
\\[0.5mm]
$4.5$
& $ 3.661 \times 10^{-7}$
& $ 0.164$
& $ 0.520$
\\[0.5mm]
$5.0$
& $ 1.406 \times 10^{-7}$
& $ 0.084$
& $ 0.267$
\\ \hline
\label{Tab:sig_with}
\end{tabular} 
\end{center}
\end{table}

\begin{table}[h]
\begin{center}
\caption{Significance of the $A_{1}^{\ 1}$ discovery on 100~TeV 
LHC run, $S/\sqrt{S+B}$, for integrated luminosity $\mathcal{L} 
= 300~\text{fb}^{-1}$ and $3000~\text{fb}^{-1}$. Here $S$ 
and $B$ are numbers of signal event and background event, 
respectively.}
\vspace{3mm}
\begin{tabular}{llllll}
\hline
$M_{11}$~[TeV]
& $\sigma_\text{BG}$~[pb] \hspace{15mm}
& $\frac{S}{\sqrt{S+B}} \ (\text{for }300~\text{fb}^{-1})$ 
& $\frac{S}{\sqrt{S+B}} \ (\text{for }3000~\text{fb}^{-1})$
\\ \hline \hline
$2.0$
& $ 2.818 \times 10^{-3}$
& $ 182.0$
& $ 575.5$
\\[0.5mm]
$4.0$
& $ 2.497 \times 10^{-4}$
& $ 38.97$
& $ 123.2$
\\[0.5mm]
$6.0$
& $ 5.428 \times 10^{-5}$
& $ 14.01$
& $ 44.32$
\\[0.5mm]
$8.0$
& $ 1.705 \times 10^{-5}$
& $ 6.296$
& $ 19.91$
\\[0.5mm]
$10.0$
& $ 6.504 \times 10^{-6}$
& $ 3.211$
& $ 10.16$
\\[0.5mm]
$12.0$
& $ 2.811 \times 10^{-6}$
& $ 1.765$
& $ 5.580$
\\[0.5mm]
$14.0$
& $ 1.317 \times 10^{-6}$
& $ 1.027$
& $ 3.249$
\\[0.5mm]
$16.0$
& $ 6.562 \times 10^{-7}$
& $ 0.625$
& $ 1.976$
\\[0.5mm]
$18.0$
& $ 3.370 \times 10^{-7}$
& $ 0.390$
& $ 1.233$
\\[0.5mm]
$20.0$
& $ 1.801 \times 10^{-7}$
& $ 0.250$
& $ 0.789$
\\ \hline
\label{Tab:sig_100}
\end{tabular} 
\end{center}
\end{table}

We are in a position to investigate the feasibility of the $A_1^{\ 1}$
discovery at the future LHC and 100~TeV collider experiments. 
Figures~\ref{Fig:14TeV_with} and \ref{Fig:100TeV_with} show dielectron 
production cross section via $A_1^{\ 1}$ at $\sqrt{s} = 14~\text{TeV}$
and at $\sqrt{s} = 100~\text{TeV}$.
Solid curve represents the total cross section, and other curves show
partial cross sections for the subprocesses with large contributions.
Shaded region is excluded by the direct search at the
$\sqrt{s}=8~\text{TeV}$ run (see Fig.~\ref{Fig:8TeV_with}).

Here we briefly see each contribution to the dielectron production.
The largest contribution almost throughout the mass region we consider
comes from $b \bar b \to A_1^{\ 1} \to e^+ e^-$, nonetheless $b$- and
$\bar{b}$-quark distributions in a proton are very small.
This is because there is no suppression from the off-diagonal elements
of the CKM matrix.
Similarly the process $d \bar b \to A_1^{\ 1} \to e^+ e^-$ also gives
large contribution.
Compared with the process $b \bar b \to A_1^{\ 1} \to e^+ e^-$, the
cross section of this process gets suppression by
$|(V_\text{CKM})_{td}|^2$, but enhancement by large distribution of 
$d$-quark in a proton. 
Other processes have small contributions due to both the suppressions 
from the CKM off-diagonal elements and the low densities of sea quarks
in a proton.

We quantitatively discuss the feasibility of the $A_1^{\ 1}$ discovery.
As an indicator of the $A_1^{\ 1}$ discovery reach, we evaluate the
significance $S/\sqrt{S+B}$ (Table~\ref{Tab:sig_with} for the 14~TeV LHC
run and Table~\ref{Tab:sig_100} for the 100~TeV LHC run).
Here $S$ and $B$ are the numbers of the dielectron signal and its
SM background, respectively.
In the evaluation of the significance, we take $300~\text{fb}^{-1}$ and 
$3000~\text{fb}^{-1}$ as an integrated luminosity.
Based on the discussion in Ref.~\cite{Aad:2014cka}, we take the product
of the acceptance and efficiency to be 0.6 for each point.

We include only SM Drell-Yan production, $pp \to Z^*/\gamma^* 
\to e^+ e^-$, in the evaluation of the SM background, because this is
the dominant contribution in the region of $m_{ee} \gtrsim 2~\text{TeV}$
(see TABLE~V in Ref.~\cite{Aad:2014cka}).
The background cross section is calculated per an energy bin of $3 \times 
\Gamma_{A_{11}}$ centered at each $M_{11}$.

Three sigma significance is an important milestone, which implies an
``evidence''. 
Table~\ref{Tab:sig_with} shows that, with the integrated luminosity of 
$3000~\text{fb}^{-1}$ at $\sqrt{s} = 14~\text{TeV}$, the family gauge 
boson with $M_{11} \lesssim 3.2~\text{TeV}$ can be identified as an
``evidence'', and assists the model to be confirmed. 
Similarly, Table~\ref{Tab:sig_100} shows that, at $\sqrt{s} =
100~\text{TeV}$, the family gauge boson with $M_{11} \lesssim
14~\text{TeV}$ is within the reach of the ``evidence''.

One of the key ingredients for the discrimination between our scenario
and other scenarios is to check the decay properties of $A_1^{\ 1}$.
Since $A_1^{\ 1}$ couples with electron but not with muon and tau
lepton, the discovery of dielectron resonance in the absence of dimuon
and ditau resonances at the same mass suggests the existence of $A_1^{\
1}$.
Such an unequal rate in the dilepton signal is one of the clear
signatures of the model, but not yet a sufficient evidence.
We have to check the other specific features of $A_1^{\ 1}$: (i) unequal
rates of diquark resonance, i.e., $\text{Br}(A_1^{\ 1} \to b \bar b) \gg
\text{Br}(A_1^{\ 1} \to \text{light flavors})$, e.g., 
(i\hspace{-1pt}i) the ratio between branching ratios of $e^+ e^-$ 
and $b \bar b$ final states, $\text{Br}(A_1^{\ 1} \to b \bar b)/\text{Br} 
(A_1^{\ 1} \to e^+ e^-) \simeq 3$.
(i\hspace{-1pt}i\hspace{-1pt}i) confirmation of the spin of $A_1^{\ 1}$ 
by the angular analysis in the dielectron events.

Another key ingredient for the discrimination and the confirmation of
the model is the cross sections of $b \bar{b} \ (t \bar{t})$ associated
$A_{1}^{\ 1}$ productions.
The FGB $A_{1}^{\ 1}$ interacts with top and bottom quarks, but not with
other quarks when we omit the intergenerational mixing.
Thus, the cross sections of $A_{1}^{\ 1}$ production associated with $b
\bar{b} \ (t \bar{t})$ must be larger than 
that of $A_{1}^{\ 1}$ production associated with light flavor jets.
Thus the measurement of $\sigma(pp \to A_{1}^{\ 1}+ b \bar{b} \, (t
\bar{t}))$ is a nice complementary check of the scenario.
Figures~\ref{Fig:14TeV_1body_bb_tt} and \ref{Fig:100TeV_1body_bb_tt} 
show the cross sections of the processes $pp \to A_{1}^{\ 1}$ and $pp
\to A_{1}^{\ 1}+ b \bar{b} \, (t \bar{t})$ at $\sqrt{s}=14~\text{TeV}$ 
and at $\sqrt{s}=100~\text{TeV}$, respectively. 
The events for $b \bar{b}$ associated $A_1^{\ 1}$ production can be
safely distinguished from those for the inclusive $A_1^{\ 1}$ production
via $b\bar b$ annihilation by requiring the $b$-tagged jets with large
transverse momenta.
Thus, for the estimation of the cross section, we impose a cut on each
$b$-jet:
(i) large transverse momentum $p_{T}^{b(\bar{b})} > 25~\text{GeV}$ 
(i\hspace{-1pt}i) pseudo-rapidity smaller than $|\eta| < 2.5$.
Here, the cut (i\hspace{-1pt}i) is required by the detector coverage for
$b$ tagging.
Although we have to perform a simulation study for detailed analysis, 
we expect the complementary check can be available 
for $A_1^{\ 1}$ with the mass up to several TeV.

Finally, we consider an alternative case that $U_{u} = (V_\text{CKM})^
\dagger$ and $U_{d} = \boldsymbol{1}$, and show the feasibility of 
discrimination of these two cases at the LHC.
Figure~\ref{Fig:ratio} shows the ratio of total cross sections of
dielectron production via $A_{1}^{\ 1}$ in each case,
$\sigma_\text{total} (U_{u} = (V_\text{CKM})^\dagger) /
\sigma_\text{total} (U_{d}=V_\text{CKM})$.
The deviation from unity in the ratio comes from the difference of 
the contributions from the off-diagonal components of $U_{u}$ and
$U_{d}$.
Figures~\ref{Fig:14TeV_with} and \ref{Fig:100TeV_with} show that, 
in the case of $U_{u} = \boldsymbol{1}$ and $U_{d} = V_\text{CKM}$, 
a subprocess with the initial state of $d+\bar{b}$ sizably contributes
to the total cross section. 
On the other hand, in the alternative case of $U_{u} = (V_\text{CKM})^
\dagger$ and $U_{d} = \boldsymbol{1}$, due to the tiny distributions of
$t$- and $\bar{t}$-quarks in a proton, the subprocesses with the initial
states of $u+\bar{t}$, $\bar{u}+t$, $c+\bar{t}$, and so on give negligible 
contributions.
Thus the ratio of total cross sections is estimated to be
\begin{equation*}
\begin{split}
   \frac{\sigma_\text{total} (U_{u} = (V_\text{CKM})^\dagger)}
   {\sigma_\text{total} (U_{d}=V_\text{CKM})} 
   \simeq 
   \frac{\sigma(b \bar{b} \to e^{+} e^{-})}
   {\sigma(b \bar{b} \to e^{+} e^{-}) + \sigma(d \bar{b} \to e^{+} e^{-})}, 
\end{split}      
\tag{3.5}
\end{equation*}
which is significantly smaller than unity. 
Thus, the measurement of the dielectron cross section can discriminate
the two cases.
In addition the signals of $A_{1}^{\ 1}$ flavor violating decays, 
e.g., $A_{1}^{\ 1} \to u \bar t \hspace{1mm} (\bar u t)$ and 
$A_{1}^{\ 1} \to c \bar t \hspace{1mm} (\bar c t)$, yield the 
information of $U_{u}$. 
The same statement is applied for the down-type quarks and $U_d$.
To study the structure of quark mixing matrices, we need more dedicated
analyze on the events with more complicated hadronic final states, which
is beyond the scope of the paper.

\vspace{5mm}

\noindent{\large\bf 4 \ Concluding remarks}
 
In the U(3) family gauge model with twisted family number assignment
(Model B)~\cite{Koide_PLB14}, the FGB $A_1^{\ 1}$ couples to the first
generation leptons, while it does to the third generation quarks.
The lowest FGB $A_1^{\ 1}$ can take a considerably smaller mass, 
for example, of an order of a few TeV, 
compared with the conventional FGB models. 
The direct measurement of $A_{1}^{\ 1}$ at collider experiments
is one of the most convincing evidence for the models with family gauge
symmetry.
In this paper we have argued that the most clear observable is the
dielectron signal.
We evaluated the production cross section and the significance of the
signal for the 14~TeV and 100~TeV LHC run. 
At the 14~TeV and 100~TeV LHC run with the integrated luminosity ${\cal
L}=3000~\text{fb}^{-1}$, the FGB with $M_{11} \lesssim 3.2~\text{TeV}$ 
and $M_{11} \lesssim 14~\text{TeV}$ is within the reach of the
``evidence'', respectively (see Tables~1 and 2). 
In order to confirm or rule out the model, a key ingredient is to check
the characteristic interactions of the FGB with leptons and quarks
[Eq.~(2.1)].  
We have evaluated the cross sections of $b \bar{b}$ ($t \bar{t}$)
associated $A_1^{\ 1}$ production, which is useful for the complementary 
check of the interactions with leptons and quarks. 
If we observe a $e^+ e^-$ peak at the LHC experiments,
whether it is really a FGB or not can be checked by
searching for the similar peak in $\mu^+ \mu^-$ and also $\tau^+\tau^-$
modes at the same invariant mass.
Measurement of the branching ratio $Br(A_1^{\ 1} \rightarrow b
\bar{b})$ also plays an essential role in identifying whether it is
really the FGB $A_1^{\ 1}$ or not.

Finally, we would like to mention the discrimination of the family
number assignment for quarks.
We have two types of the assignment (see Eq.~(2.4)).
The different assignment gives rise to the different mass of the next
lightest FGB (see Table 1 in Ref.~\cite{Koide_PLB14}), and its decay 
modes are also different.
However, because the next lightest FGB is predicted to be too heavy,
there may be difficult to find the direct evidence at the LHC.
One of the probe to the next lightest FGB is the $\mu$-$e$ conversion in
nuclei.
COMET, DeeMe and Mu2e experiments will launch soon, and search for the
$\mu$-$e$ conversion
signal~\cite{Cui:2009zz,Natori:2014yba,Bartoszek:2014mya}. 
It will be worthwhile to investigate this process to discriminate the
family number assignment in the model.

In conclusion, the most clear detection of the FGB $A_1^{\ 1}$
is to observe a $e^+ e^-$ peak at the LHC.
We are looking forward to observing such a peak in the forthcoming data
at $\sqrt{s}= 14$~TeV and at $\sqrt{s}= 100$~TeV.

\vspace{5mm}

{\Large\bf Acknowledgments} 

One of the authors (YK) thanks T.~Yamashita  for helpful conversations.
and Y. Sumino for valuable comments on the U(3) familly symmety
beraking scale.@
He (YK) also thanks H.~Terazawa and I.~Sogami for informing him 
helpful comments on works on family symmetries in earlier stage.
M.Y.\ thanks K.~Tobe for useful discussions. 
This work was supported in part by the Grant-in-Aid for 
the Ministry of Education,
Culture, Sports, Science, and Technology, Government of Japan,
No.\ 25003345 (M.Y.).
The work of H.Y.\ was supported by JSPS KAKENHI Grant Number 15K17642.
  

 \vspace{10mm}

%

%

\end{document}